\documentclass[]{article}
\usepackage{zeldovich}

\usepackage{cite} 
\usepackage{amssymb} 
\usepackage{amsfonts} 
\usepackage{amsmath} 
\usepackage{breqn}
\usepackage{slashed} 

\newcommand{\beq}{\begin{equation}}
\newcommand{\eeq}{\end{equation}}
\newcommand{\beqa}{\begin{eqnarray}}
\newcommand{\eeqa}{\end{eqnarray}}

\newcommand{\half}{\frac{1}{2}}

\newcommand{\pbr}[2]{ \{ \hspace*{-2.6pt} [ #1 , #2\hspace*{1.4 pt} ] 
\hspace*{-2.6pt} \} }

\newcommand{\we}{\wedge}

\newcommand{\xx}[1]{\raisebox{1pt}{$\stackrel{#1}{X}$}}

\newcommand{\ff}[1]{\raisebox{1pt}{$\stackrel{#1}{F}$}}

\newcommand{\der}{\partial}

\newcommand{\inn}{\hspace*{2pt}\raisebox{-1pt}{\rule{6pt}{.3pt}\hspace*
{0pt}\rule{.3pt}{8pt}\hspace*{3pt}}}

\newcommand{\ka}{\varkappa}

\newcommand{\Psib}{\overline{\Psi}}
\newcommand{\Phib}{\overline{\Phi}}

\newcommand{\what}[1]{\widehat{#1}}

\newcommand{\bx}{{\mathbf{x}}}

\newcommand{\BPsi}{{\bf \Psi}} 
   
\newcommand{\hP}{\hat{P}}

\DeclareMathOperator{\Tr}{Tr} 

\begin{document}

\title{\bf On precanonical quantization of gravity}
\author{I.V. Kanatchikov\\
\small\itshape 
National Quantum Information  Center in Gda\'nsk,  
\small 81-824 Sopot, Poland } 
\date{} 
\maketitle

\begin{abstract}
Precanonical quantization 
is based on the mathematical structures of the De Donder-Weyl Hamiltonization of field theories.
The resulting formulation of quantum gravity describes the quantum geometry of space-time 
 in terms of operator-valued distances and 
the transition amplitudes between the values of spin connection 
at different points of space-time, which obey the covariant 
precanonical analogue of the Schr\"odinger equation. 
 In the context of quantum cosmology 
  the theory predicts 
a probability distribution of a cosmological spin-connection field, 
which may have an observable impact on 
the large scale structures in the universe. 
\end{abstract}

\paragraph{Introduction.}  
The attempts to construct quantum theory of gravity using 
the methods of QFT originating from  canonical quantization 
in Minskowski space-time are known to lead to certain 
technical and conceptual difficulties. One of them is the 
so-called ``problem of time" which can be traced back to the 
distinguished role of time in the canonical Hamiltonian 
formalism. The approach of precanonical quantization is 
based on a different Hamiltonization in field theory, which 
does not distinguish between the space and time variables. 
The space-time variables are treated on the equal footing 
as a multidimensional 
 analogue 
of the time parameter in mechanics.  This Hamiltonization is known 
in the calculus of variations 
as the De Donder-Weyl  (DW) theory (see e.g. [1]).\vspace*{-5pt} 

\paragraph{DW Hamiltonization.} 
For a Lagrangian density {\small $L = L(y^a, y^a_\mu, x^\nu)$,}  
which is a function of the fields variables $y^a$, 
their first space-time derivatives {\small $y^a_\mu$,} 
and the space-time variables {\small$x^\mu$,} one defines the {\em polymomenta}:
{\small $p^\mu_a := \frac{\der L}{\der y^a_\mu} ,$}
and the {\em DW Hamiltonian function}: 
{\small $H(y^a, p^\mu_a, x^\mu) := y^a_\mu(y,p) p^\mu_a - L .$} 
Then, in the regular case {\small $\det (\der^2L/\der y^a_\mu \der y^b_\nu)\neq 0$,}
the Euler-Lagrange field equation can be written in the {\em DW Hamiltonian form}: 
\vspace*{-8pt}
{\small \beq \label{dw}
\der_\mu y^a (x) = {\der H}/{\der p^\mu_a} , \quad  
\der_\mu p^\mu_a (x) = - {\der H}/{\der y^a}   , \vspace*{-8pt} 
\eeq}
which requires neither a splitting into the space and time nor infinite-dimensional
spaces of field configurations.  
Here the analogue of the extended configuration
space is~the~space~of~field variables~$y^a$~and 
space-time variables $x^\mu$, and the analogue of the extended phase space is a finite dimensional space of 
$p^\mu_a, y^a$~and~$x^\mu$. Classical fields are sections in the corresponding bundles over the space-time.\vspace*{-5pt}  

\paragraph{\bf DW theory and precanonical quantization.}  
Field quantization based on the above Hamiltonization uses the mathematical structures of DW Hamiltonian formalism which were found in our earlier papers [2]. 
The {\em polysymplectic form } on the polymomentum phase space: 
{\small $\Omega := dp^\mu_a \we dy^a \we \varpi_\mu,$} where \
{\small$\varpi_\mu:=\der_\mu \inn\varpi$}   
 and $ \varpi\!\!:=$\mbox{\small $dx^1\we...\we dx^n$} is the volume form on $n$-dimensional space-time, 
leads to the definition of Poisson brackets on  forms of 
different degrees $p$ and $q$ which represent dynamical variables: 
 {\small $\pbr{\ff{p}_1}{\ff{q}_2} = (-)^{(n-p)} \xx{n-p}_1\inn\,d\ff{q}_2,$} 
where {\small$\xx{n-p}$} is a Hamiltonian multivector field related to the $p$-form {\small$\ff{p}$} 
via the map: 
{\small $\xx{n-p}\inn\,\Omega = d \ff{p},$}   $p=0,1,...,(n-1)$. The space of forms for which this map exists is closed 
with respect to the $\bullet$-product: 
{\small $\ff{p}\bullet \ff{q}:= *^{-1} (*\ff{p}\we*\ff{q}),$} and the bracket operation equips it with the structure of the {\em Gerstenhaber  algebra}, 
which 
appears here as a generalization of the Poisson algebra structure to the 
DW Hamiltonian formulation. 
Precanonical quantization relies on the fundamental brackets [2,3]:\vspace*{-1pt} 
{\small 
$$
\pbr{p_a^\mu\varpi_\mu}{y^b}
= 
\delta^b_a , \;
\pbr{p_a^\mu\varpi_\mu}{y^b\varpi_\nu}
=
\delta^b_a\varpi_\nu, \;
\pbr{p_a^\mu}{y^b\varpi_\nu}
= 
\delta^b_a\delta^\mu_\nu .\vspace*{-3pt}
\refstepcounter{equation} 
\eqno {  (\theequation a,b,c) } \label{fundbr}
$$ 
}Their quantization leads to the representation  of polymomenta 
and {\small $(n-1)$}-forms $\varpi_\mu$ as Clifford-valued operators [3]:\vspace*{-4pt}
{\small  $$
\hat{p}{}^\nu_a = -i \hbar \ka \gamma^\nu \frac{\der}{ \der y^a}, 
\quad 
 \what{\varpi}_\nu = \frac{1}{\ka}\gamma_\nu,\vspace*{-3pt}
\refstepcounter{equation} 
\eqno {  (\theequation a,b) } \label{pmuop}
$$
}where the parameter {\small$\frac{1}{\ka}$} appears on the dimensional grounds as a very small 
quantity 
of the dimension of {\small $(n-1)$}-volume; \mbox{one could dub it a quantum of space.}

The precanonical analogue of the 
Schr\"odinger equation [3,4]: 
{\small \beq \label{nse}
i\hbar\varkappa\gamma^\mu\der_\mu \Psi = \what{H}\Psi , 
\eeq 
}where {\small $\what{H}$} is the operator of  DW Hamiltonian and 
{\small $\Psi(y^a,x^\mu)$} is a Clifford-valued wave function,  
is suggested by the fact that the DW Hamiltonian equations~can be writen in terms of the bracket 
of the fundamental variables in (\ref{fundbr})  with~$H$~[2], 
which will generate their total co-exterior differential [3,4]
(similarly to the generation of the total time derivative by the Poisson bracket with the Hamilton's function in mechanics). 
We can also argue [3] that (\ref{nse}) allows us to \mbox{obtain} 
the classical field equations in DW form as 
the equations for the \mbox{expectation} values of the corresponding precanonical operators, and to reproduce  the \mbox{Hamilton-} Jacobi  equation  
of  DW~theory~{\small [1] }in the classical limit.  The~scalar~product~is~re-
\newpage \hspace*{-15pt}lated to the conservation law of (4): 
{\small $\der_\mu 
\int\!dy  \Tr \left[ \Psib \gamma^\mu \Psi \right]  =0 ,$} where 
{\small $\Psib:=\gamma^0 \Psi^\dagger\gamma^0.$} 

When applied to the scalar field theory [3] with {\small $L= \half \der_\mu y \der^\mu y - V(y),$} 
we obtain 
 {\small $\hat{H} = -\frac{1}{2}\hbar^2\ka^2  \frac{\der^2}{\der y^2} + V(y). $} 
For  the free field theory with {\small $V(y) =\frac{1}{2} \frac{m^2}{\hbar^2} y^2$} the spectrum of normal ordered {\small $\frac{1}{\ka}\hat{H}$} reproduces the mass spectrum of free particles: $mN$, where  $N$ is the quantum number of $\hat{H}$.  By writing (4) in the form {\small $i\hbar\der_\mu \Psi = \hat{P}_\mu \Psi$} 
and defining {\small $\hat{y}(x) := e^{i\hP_\nu x^\nu} y\, e^{-i\hP_\nu x^\nu},$} 
we can derive the  
standard  correlators of $\hat{y}(x)$ from the precanonical theory{\,\small [5]}.\vspace*{-5pt} 

%
%

\paragraph{Standard QFT as a limiting case.} The 
comparison of probabilistic interpretations 
 of precanonical 
$\Psi(y,x)$ and the canonical Schr\"odinger wave functional 
$\BPsi([y(\bx)],t)$, 
and the corresponding equations,
 allows us to establish 
a relation between 
 them{\,\small [6]\,}in terms of the Volterra's 
multidimensional product integral{\,\small  [7]:}
\beq \label{schr-prod}
\BPsi = {\sf Tr}\left \{\prod_\bx 
e^{-iy(\bx)\alpha^i\der_iy(\bx) d\bx} 
  \Psi_\Sigma (y(\bx), \bx, t)_{\mbox{\large $\rvert$} \scriptscriptstyle    
  \frac{1}{\ka} \beta \mapsto d\bx }
\right \} , 
\eeq 
where {\small $\Psi_\Sigma (y(\bx), \bx, t)$} is the restriction of {\small$\Psi(y,x)$} to the subspace 
{\small $\Sigma$: \!$(y\!=\!y(\bx), x^0\!=\!t),$} and the notation $\Psi_\Sigma{}_{\mbox{$\rvert$}\scriptscriptstyle  \frac{1}{\ka} \beta \mapsto d\bx }$ means that every $\beta/\ka$ in the expression of $\Psi$ is replaced by $d\bx$ 
before the product integral is evaluated. 
In [6b]
it is expli\-citly 
demonstrated  how this product integral formula 
leads to the vacuum state wave functional of free scalar field from the ground state solution of 
precanonical Schr\"odinger equation. Formula (\ref{schr-prod}) also tells us that the 
standard QFT obtained from canonical quantization is a limiting case of vanishing $\frac{1}{\ka}$ of the 
theory obtained from precanonical quantization. To be more precise, the limiting transition involves 
the inverse 
 of the 
quantization map in (3b)  at $\nu=0$: ${\frac{\beta}{\ka}\mapsto d\bx}$, that implies 
an infinitesimal quantum of space $\frac{1}{\ka}$.\vspace*{-5pt}  


\paragraph{Precanonical quantization of gravity.} 
While precanonical quantization of metric gravity was discussed by us earlier [8], 
the appearance of the Dirac operator in (4) makes the vielbein formulation of general relativity a preferable starting point for precanonical quantization. Here the Lagrangian density \vspace*{-2pt}
{\small  \beq \label{lagr}
{\mathfrak L}=  \mbox{$\ \mbox{$\frac{1}{\kappa_E}$}$} {\mathfrak e} e^{[\alpha}_I e^{\beta ]}_J 
\left(\der_\alpha \omega_\beta{}^{IJ} +\omega_\alpha {}^{IK}\omega_{\beta K}{}^J\right) + \mbox{$\mbox{$\frac{1}{\kappa_E}$}$}\Lambda {\mathfrak e} \vspace*{-1pt}
\eeq 
}with  the vielbein components $e^\mu_I$,  the torsion-free spin-connection coefficients $\omega_\alpha^{IJ}$, 
 the Einstein's gravitational constant $\kappa_E:= 8\pi G$,  and ${\mathfrak e}:= \det{||e_\mu^I}||$ 
leads to the singular DW Hamiltonization with the primary constraints 
{\small 
\beq{\mathfrak p}{}^\alpha_{e^I_\beta}:=
\frac{\der {\mathfrak L} }{\der_\alpha e^I_\beta}  \approx 0 , 
\;\;  \mathrm{} \;\;  
{\mathfrak p}{}^\alpha_{\omega_\beta^{IJ}} :=\frac{\der {\mathfrak L} }{\der_\alpha{\omega_\beta^{IJ}}}
\approx 
\mbox{$ \mbox{$\frac{1}{\kappa_E}$}$}
{\mathfrak e} e^{[\alpha}_Ie^{\beta ]}_{J } .\vspace*{-2pt}
\eeq 
}
We use our generalization of Dirac's approach to constrained systems and the 
\newpage 
\hspace*{-15pt}Dirac bracket to singular DW theories [9]. 
The Poisson 
brackets of {\small $(n-1)$}-forms 
constructed from the constraints: 
{\small $
\mathfrak{C}_{e_\beta^I}
:={\mathfrak p}_{e_\beta^I}^\alpha\varpi_\alpha, 
\; 
\mathfrak{C}_{\omega_\beta^{IJ}}
:=  
{\mathfrak p}{}^\alpha_{\omega_\beta^{IJ}} \varpi_\alpha
-
\mbox{$\ \mbox{$\frac{1}{\kappa_E}$}$}
{\mathfrak e} e^{[\alpha}_Ie^{\beta ]}_{J } \varpi_\alpha 
$}:\vspace*{-2.5pt}
 {\small 
\beq \label{cbr}
 \pbr{\mathfrak{C}_e}{\mathfrak{C}_{e'}} =0 , \; 
 \pbr{\mathfrak{C}_\omega}{\mathfrak{C}_{\omega'}} =0 ,\;
 \pbr{\mathfrak{C}_{e_\gamma^K}}{\mathfrak{C}_{\omega_\beta^{IJ}}} 
 = - \frac{1}{\kappa_E}\frac{\der}{\der{e_\gamma^K}}
 \left( 
 {\mathfrak e} e^{[\alpha}_Ie^{\beta ]}_{J } 
 \right)  \varpi_\alpha  , \vspace*{-2.5pt}
\eeq
}indicate that the primary constraints of DW formulation are {\bf second class}. Using our generalization 
of the Dirac bracket to DW theory 
we were able to show {\small [10]} that the Dirac brackets between the vielbeins and their polymomenta 
vanish, e.g. 
{\small $\pbr{{\mathfrak p}^\alpha_e \varpi_\alpha}{e'}{\!}^D=0,$ } and the Dirac brackets between the 
 spin connection coefficients and their polymomenta are the same as if there were no constraints, e.g. 
{\small $\pbr{{\mathfrak p}^\alpha_\omega \varpi_\alpha}{\omega'}{\!}^D 
= \delta_\omega^{\omega'}.$ } This fact 
 simplifies quantization performed in{\small \,[10]\,}using the generalized Dirac's quantization rule:  
 {\small $[\hat{A}, \hat{B}]= 
- i\hbar \what{\mathfrak{e}\pbr{A}{\!B\!}{\!}}{}^D, 
$}
where the operator  of $\mathfrak{e}$ 
ensures that tensor densities are  quantized as density-valued operators.

From quantization of fundamental Dirac brackets and using the \mbox{equations of} 
constraints (7) we conclude that the precanonical wave function does not depend on vielbein variables, 
i.e. {\small $\!\!\Psi=\Psi(\omega^{IJ}_\alpha, x^\mu)$}, and obtain a represenation of 
the opera\-tors of vielbeins: 
{\small $\hat{e}{}^\beta_I = -i \hbar\ka\kappa_E \bar{\gamma}^{J}\frac{\der}{\der \omega_{\beta}^{IJ}}$}, 
and the polymomenta of spin-connection: 
{\small $\hat{{\mathfrak p}}{}^{\alpha}_{\omega_\beta^{IJ}} 
 = - \hbar^2\ka^2\kappa_{{\!}_E}
\,\hat{\mathfrak e}\,
\bar{\gamma}^{KL}\frac{\der}{\der \omega_{[\alpha}^{KL}}
\frac{\der}{\der \omega_{\beta]}^{IJ}}$},   where 
$\bar{\gamma}^{J}$ 
are the fiducial Minkowskian Dirac matrices 
 and
{\small $\hat{\mathfrak{e}}{} = 
\left( \frac{1}{n!} \epsilon^{I_1...I_n}\epsilon_{\mu_1...\mu_n} 
\hat{e}{}^{\mu_1}_{I_1} ... \hat{e}{}^{\mu_n}_{I_n} \right)^{-1}$}.
This allows us to construct the operator of DW Hamiltonian density ${\mathfrak e}H$ restricted to the 
constraints surface $C$: {\small ${({\mathfrak e} H)|_C} = 
- {\mathfrak p}{}^{\alpha}_{\omega_\beta^{IJ}} \omega_\alpha^{IK} \omega_{\beta K}{}^{J} 
- \frac{1}{\kappa_E} \Lambda{\mathfrak e}$}, which is derived from (6), 
so that\vspace*{-3pt}{\small 
\begin{equation} \label{hgrop}
\what{H} = \hbar{}^2\ka^2\kappa_E \bar{\gamma}^{IJ} 
\omega_{[\alpha}{}^{KM}\omega_{\beta] M}{}^L 
\frac{\der}{\der \omega_{\alpha}^{IJ}} \frac{\der}{\der \omega_{\beta}^{KL}} 
- \frac{1}{\kappa_E} \Lambda , \vspace*{-2pt}
\end{equation}
}and 
to obtain the covariant 
precanonical analogue of the Schr\"odinger equation for quantum gravity: 
 {\small 
\beq \label{nsepsi}
i \hbar\ka 
\what{\slashed\nabla}  \Psi = 
\what{H} \hspace*{-0.0em} \Psi ,   
\eeq
}where 
 {\small $\what{\slashed\nabla} 
:=  \widehat{\gamma}{}^\mu(\der_\mu+   
 \frac{1}{4} \omega_{\mu IJ} \bar{\gamma}^{IJ})$,}   
in the explicit form:  \vspace*{-1pt}
{\small 
\begin{dmath} \label{wdw}
\bar{\gamma}{}^{IJ} 
 \left( \der_\mu +   \frac{1}{4} \omega_{\mu KL}\bar{\gamma}^{KL} 
  - 
  \omega_{\mu M}^{K}\omega_{\beta}^{ML} 
  \frac{\der}{\der \omega_{\beta}^{KL}} 
  \right) 
\frac{\der}{\der \omega_{\mu}^{IJ}}   
   \Psi 
       + 
       \lambda \Psi =0, 
\end{dmath}\vspace*{-1pt}
}
{\hspace*{-19pt}} 
where  $\lambda:= {\Lambda}/({\hbar^2\ka^2\kappa_E^2})$ is a dimensionless constant. 

The Hilbert space of the theory is defined by the scalar product with the operator-valued invariant 
measure on the space of spin-connection coefficients: {\vspace*{-0pt}} 
{\small
\beq 
\left\langle \Phi | \Psi \right\rangle 
:=  \Tr \int  \Phib \, \what{[d\omega]}_{} \Psi, \quad 
\what{[d\omega]}=\hat{{\mathfrak e}}{}^{- n(n-1)}\prod_{\mu, I<J} d \omega_\mu^{IJ}, 
\vspace*{-3pt}
\eeq}which is obtained using the arguments similar to 
 those in [11]. 
It is interesting to note that the normalizability of precanonical wave functions actually implies 
the quantum singularity avoidance, because $\Psi$ should vanish at large $\omega$-s, i.e. 
at large space-time curvatures.\newpage 

Note that the potential  issues related to the indefiniteness of {\small$\Tr [\Psib\Psi]$}  
and the 
gauge fixing, i.e. the choice of the coordinate systems 
and local 
orientations 
of vielbeins on the average, when extracting a physical information from the solutions of 
(11),  are not yet sufficiently clarified. 

The Green functions of (\ref{wdw}):  
{\small $\langle \omega,x |\omega',x' \rangle$,} which are the transition amplitudes from the 
values of the spin-connection components  $\omega'$ at the point $x'$ to the values $\omega$ at the point $x$,  provide an inherently  quantum description of 
space-time geometry, which generalizes  the classical description of  geometry in terms of smooth spin-connection fields $\omega(x)$.  
Besides, the distances between points are given by quantum operators, because the metric tensor in the present formulation is operator-valued: 
{\small$\what{g^{\mu\nu}} = -\hbar^2 \ka^2\kappa_E^2 
\eta^{IJ}\eta^{KL}\frac{\der^2}{\der\omega^{IK}_\mu \der\omega^{JL}_\nu}$.}
This type of description of quantum geometry of space-time in terms of 
   the 
transition amplitudes on the connection bundle and 
   the 
operator-valued 
 metric structure on the space-time 
complements the current intuitive ideas about the quantum space-time   
suggested by quantum geometrodynamics, loop quantum gravity, string theory  and non-commutative geometry. 

The fact that all dimensionful constants in (11) are absorbed in one dimensionless constant 
{\small$\lambda$,} 
which depends on the ordering of operators $\omega$ and $\der_\omega$,  
seems to be important. Knowing {\small $\lambda$} we would be able to determine the value of our constant $\ka$. A naive estimation yields {\small $\lambda \sim n^6$} and then $\ka$ at $n=4$ is at the nuclear scale, which is unexpected. If, however, we assume that  $\ka$  is Planckian, then the estimated value of 
$\Lambda$ is $\sim \!10^{120}$ times higher  than the observable one, 
which is a usual problem in naive QFT-based estimations of $\Lambda$.  This coincidence confirms  that  
$\ka$ of precanonical quantization is related to the ultra-violet cutoff scale in standard QFT and 
indicates 
that the cosmological constant is not likely to be related to the ground state of pure quantum gravity 
alone.\vspace*{-5pt}

\paragraph{\bf Precanonical quantum cosmology.} 
For {\small $n\!=\!4$ flat FLRW} metric with a  harmonic time  $\tau$:\vspace*{-8pt}{\small 
\beq \label{flrw}
\;\;ds^2=a(\tau)^6d\tau^2 - a(\tau)^2 d\bx^2 , \vspace*{-2pt}
\eeq}let us choose 
$e^0_\nu = a^3 \delta^0_\nu$, $e^I_\nu = a \delta^I_\nu$,    
%
so that $\omega_\nu^{I0} ={\dot{a}}/{2a^3} \delta^I_\nu=:\omega\delta^I_\nu$ 
($I=1,2,3$). 
Then the precanonical Schr\"odinger equation, eq.~(\ref{wdw}),  takes the form \vspace*{-6pt}{\small
\beq \label{wdww}
\Big ( 2 \sum_{i=I=1}^3 
   \gamma^{0I}\der_\omega\der_i +
 3\omega\der_\omega + 
 c \Big )\Psi=0,  \vspace*{-5pt}
\eeq}where,  if\ $\omega\der_\omega$\ is Weyl-ordered, 
{\small$c= \frac{3}{2} + \frac{\Lambda}{({\hslash\ka\kappa_E})^2}$}   
 is the effective cosmological constant. 
By separation of  variables {\small $\Psi := u(\bx)f(\omega)$} we obtain: 
\mbox{\small
$2 \sum_{i=I=1}^3
\gamma^{0I} \der_i u = iq u,$} and {\small\; $(iq\der_\omega + 3\omega\der_\omega + c ) f=0 .$ 
}
The solution of the latter: 
 {\small 
$ f\sim (3\omega+iq)^{-c/3},  $
}yields the probability density  
{\small \beq \label{distr}
\rho(\omega) := \bar{f}f \sim \frac{1}{(9\omega^2 + \bar{q}q)^{c/3}}  .  
\eeq}One can either interpret it as a distribution of quantum universes 
according to the value of $\omega = {\dot{a}}/{2a^3}$, i.e. essentially the expansion rate $\dot{a}$, 
or as a spatially homogeneous distribution function of quantum fluctuations of 
the random cosmological spin-connection  field $\omega$. The possibility of the latter point of view  
within the precanonical approach 
makes the usual interpretational issues of quantum cosmology much less troublesome. 

Note that our discussion is based on a  toy quantum cosmology model, where no influence of matter fields 
is taken into account so far. It would be interesting to investigate if the probability distribution 
of spin connection (15) predicted by precanonical quantum gravity theory manifests itself 
in the large scale structures in the universe and can be tested 
by cosmological observations.\vspace*{-5pt}

\small

\paragraph{\small\bf Acknowledgement.} {\small I thank Prof. Yu. Kurochkin and the organizing committee of Zeldovich-100  in Minsk for the opportunity to present this talk and enjoy the  meeting.}\vspace*{-5pt}

\paragraph{\bf References \\} 

{\hspace*{-9pt}}[1] H. Kastrup, 
{\em Phys. Rep. } {\bf 101} (1983) 1.


{\hspace*{-15pt}[2] I.V.~Kanatchikov, 
\emph{Rep. Math. Phys.} {\bf 41} (1998) 49,  
 \texttt{hep-th/9709229};  \\
{\hspace*{13pt}I.V.~Kanatchikov, 	
{\em Rep. Math. Phys.} {\bf 40} (1997) 225, 
 \texttt{hep-th/9710069}.

{\hspace*{-15pt}[3] I.V.~Kanatchikov, 
{\em AIP Conf. Proc.} {\bf 453}  (1998) 356, 
\texttt{hep-th/9811016}; \\
{\hspace*{13pt}I.V.~Kanatchikov,  
\emph{Rep. Math. Phys.} {\bf 43} (1999) 157, \texttt{hep-th/9810165}.

{\hspace*{-15pt}[4] I.V.~Kanatchikov, 
in: {\sl Differential Geometry and its Applications, } ed. O. Kowalski 
{\hspace*{13pt}e.a.
(Opava, 2001)
p. 309, 
 \texttt{hep-th/0112263}. 

{\hspace*{-15pt}[5] I.V.~Kanatchikov, Precanonical quantization: from foundations to quantum gravity, 
{\hspace*{13pt}{\em in preparation}. 

{\hspace*{-15pt}[6] I.V.~Kanatchikov, 
 {\em Adv. Theor. Math. Phys. } {\bf 18} (2014), {\tt arXiv:1112.5801};\\
{\hspace*{13pt}I.V.~Kanatchikov, 
 {\tt arXiv:1312.4518.}

{\hspace*{-15pt}[7] V.~Volterra, B.~Hostinsk\'y, 
{\sl Op\'erations Infinit\'esimales Lin\'eaires, } Gauthier-Villars, 
{\hspace*{13pt}Paris (1938).

{\hspace*{-15pt}[8] I.V.~Kanatchikov, 
\emph{Int. J. Theor. Phys.} {\bf 40}  (2001) 1121,
\texttt{gr-qc/0012074}.  \\
{\hspace*{13pt}See also:  I.V.~Kanatchikov, 
\texttt{gr-qc/9810076},  
\texttt{gr-qc/9912094}, 
\texttt{gr-qc/0004066}.

{\hspace*{-15pt}[9] I.V.~Kanatchikov,
in: {\sl Differential Geometry and its Applications, } ed. O. Kowalski 
{\hspace*{13pt}e.a.
 (World Sci., Singapore, 2008) p. 615, \texttt{arXiv:0807.3127}.

{\hspace*{-20pt}[10] I.V.~Kanatchikov, 
{\em AIP 
Conf. Proc. } {\bf 1514 } (2012) 73,  \texttt{arXiv:1212.6963}; \\
{\hspace*{13pt}I.V.~Kanatchikov, 
{\em J. \!Phys. \!Conf. \!Ser.} {\bf 442}  (2013) 012041,  {\tt arXiv:1302.2610.}

{\hspace*{-20pt}[11] C.~Misner,  {\em Rev. Mod. Phys.} {\bf 29} (1957) 497.
 

\newpage
\end{document}